\documentclass[twocolumn,prd,aps,preprintnumbers, showpacs, nofootinbib,superscriptaddress, notitlepage, 10pt]{revtex4-1}

\usepackage{enumitem}
\usepackage[colorlinks, 
linkcolor=blue, 
citecolor=blue, 
urlcolor=blue]{hyperref}
\usepackage{graphicx}
\usepackage{dcolumn}
\usepackage{bm}
\usepackage{url}
\usepackage{epsfig}
\usepackage{amsmath}
\input{epsf}
\usepackage{psfrag}
\usepackage{xcolor}
\usepackage{pdfpages}

\makeatletter
\renewcommand{\fnum@figure}{\textbf{\figurename~\thefigure}}
\renewcommand{\fnum@table}{\textbf{\tablename~\thetable}}
\makeatother

\usepackage[normalem]{ulem}
\makeatletter
\AtBeginDocument{\let\LS@rot\@undefined}
\makeatother
\newcommand{\bea}{\begin{eqnarray}}
\newcommand{\eea}{\end{eqnarray}}

\newcommand{\paren}[1]{\left( #1 \right)}

\begin{document}

\title{Physics-Informed Global Extraction of the  Universal Small-$x$ Dipole  Amplitude}
\author{Si-Wei Dai}
\email[]{swdai@mails.ccnu.edu.cn}
\affiliation{Key Laboratory of Quark and Lepton Physics (MOE) \& Institute of Particle Physics, Central China Normal University, Wuhan 430079, China}
\affiliation{Artificial Intelligence and Computational Physics Research Center, Central China Normal University, Wuhan 430079, China}

\author{Fu-Peng Li}
\email[]{fpli@fudan.edu.cn}
\affiliation{Key Laboratory of Nuclear Physics and Ion-beam Application (MOE) \& Institute of Modern Physics, Fudan University, Shanghai 200433, China}
\affiliation{Shanghai Research Center for Theoretical Nuclear Physics,
NSFC and Fudan University, Shanghai 200438, China}

\author{Long-Gang Pang}
\email[]{lgpang@ccnu.edu.cn}
\affiliation{Key Laboratory of Quark and Lepton Physics (MOE) \& Institute of Particle Physics, Central China Normal University, Wuhan 430079, China}
\affiliation{Artificial Intelligence and Computational Physics Research Center, Central China Normal University, Wuhan 430079, China}

\author{Guang-You Qin}
\email[]{guangyou.qin@ccnu.edu.cn}
\affiliation{Key Laboratory of Quark and Lepton Physics (MOE) \& Institute of Particle Physics, Central China Normal University, Wuhan 430079, China}
\affiliation{Artificial Intelligence and Computational Physics Research Center, Central China Normal University, Wuhan 430079, China}

\author{Shu-Yi Wei}
\email[]{shuyi@sdu.edu.cn}
\affiliation{Institute of Frontier and Interdisciplinary Science,
Key Laboratory of Particle Physics and Particle Irradiation (MOE), Shandong University, Qingdao, Shandong 266237, China}

\author{Han-Zhong Zhang}
\email[]{zhanghz@ccnu.edu.cn}
\affiliation{Key Laboratory of Quark and Lepton Physics (MOE) \& Institute of Particle Physics, Central China Normal University, Wuhan 430079, China}
\affiliation{Artificial Intelligence and Computational Physics Research Center, Central China Normal University, Wuhan 430079, China}

\author{Wenbin Zhao}
\email[]{WenbinZhao@ccnu.edu.cn}
\affiliation{Key Laboratory of Quark and Lepton Physics (MOE) \& Institute of Particle Physics, Central China Normal University, Wuhan 430079, China}
\affiliation{Artificial Intelligence and Computational Physics Research Center, Central China Normal University, Wuhan 430079, China}

\begin{abstract}
We extract the universal small-$x$ dipole scattering amplitude $N(r,x_B)$ from a global analysis based on a physics-informed neural network (PINN), without imposing a priori MV-type parametrization of the initial condition. The network provides a smooth and differentiable surrogate for $N(r,x_B)$, whose rapidity dependence is constrained by the collinearly improved Balitsky--Kovchegov evolution equation, while its functional form is simultaneously constrained by Deep Inelastic Scattering (DIS) data for the reduced total and charm cross sections, exclusive $J/\psi$ photoproduction measurements, and a positivity requirement for the momentum-space dipole amplitude. The resulting single universal amplitude consistently describes all fitted observables within a unified framework, alleviating the long-standing tension between total and charm channels encountered in conventional small-$x$ fits based on rigid parametric ans\"atze. Within the fitted kinematic domain, the best extracted PINN solution yields a smooth, non-negative momentum-space dipole over the full transverse-momentum range examined. Our results provide a robust and well-behaved input for Color Glass Condensate phenomenology across a broad class of high-energy processes.
\end{abstract}

\maketitle

\textbf{\textit{Introduction.}} At high energies,  Quantum Chromodynamics (QCD) enters a regime where nonlinear gluon recombination tames the growth of gluon densities, leading to gluon saturation. This regime is systematically described by the Color Glass Condensate (CGC) effective field theory~\cite{Kovchegov:2012mbw,McLerran:2002wj,Iancu:2003xm,Weigert:2005us,Gelis:2010nm,Albacete:2014fwa,Morreale:2021pnn}.
Within this framework, the small-$x$ evolution of Wilson-line correlators is governed by the JIMWLK renormalization group equation~\cite{Jalilian-Marian:1996mkd,Jalilian-Marian:1997qno,Jalilian-Marian:1997jhx,Iancu:2001md,Ferreiro:2001qy,Iancu:2001ad,Iancu:2000hn}. In the mean-field  and large-$N_c$ limit it reduces to the Balitsky--Kovchegov (BK) equation~\cite{Balitsky:1995ub,Kovchegov:1999yj}, which is a closed evolution equation for the dipole amplitude $N(r,x_B)$. This  amplitude is the universal input for a wide class of small-$x$ observables, including inclusive DIS structure functions, exclusive diffractive processes and inclusive hadron productions in $e$--$p/A$ and $p$--$p/A$ collisions.

DIS measurements, in particular of reduced total and heavy-flavor production cross sections at small-$x$ provide the clean lever arm to probe saturation dynamics. DIS off nuclei is especially compelling because saturation effects are parametrically enhanced in heavy nuclei~\cite{Kowalski:2007rw}. The future Electron-Ion Collider (EIC)~\cite{AbdulKhalek:2021gbh,Aschenauer:2017jsk,Accardi:2012qut} will enable precision measurements over a broad kinematic range in the gluon longitudinal momentum fraction $x_B$ and photon virtuality $Q^2$.
Establishing nonlinear QCD dynamics at collider energies and quantifying its impact on the small-$x_B$ structure of protons and nuclei will therefore require global analyses that simultaneously confront multiple, complementary DIS channels. In particular, combining inclusive and heavy-flavor data is essential because they probe different dipole sizes and hence different regions of the dipole amplitude. 

{
Despite substantial progress, several intertwined issues remain in small-$x_B$ global analyses. First, achieving a simultaneous description of DIS   reduced total cross section $\sigma_r$ and the charm reduced cross section $\sigma_r^{c \bar{c}}$ is  challenging~\cite{Albacete:2009fh,Albacete:2010sy,Lappi:2013zma,Mantysaari:2018zdd,Ducloue:2019jmy}. Because charm production preferentially probes smaller dipoles, $\sigma_r^{c \bar{c}}$ imposes particularly stringent constraints on the short-distance behavior of $N(r,x_B)$ and on the  conventionally assumed nonperturbative initial condition for the evolution. Second, widely used initial-condition parameterizations can yield a momentum-space dipole, $S(k_T,x_B)$ defined in Eq.~(\ref{eq:FT}),that becomes negative over parts of the $k_T$ range, which is in tension with the Fourier-positivity constraints discussed in Refs.~\cite{Albacete:2009fh,Giraud:2016lgg,Ducloue:2019jmy}. This  problem has been highlighted for some 25 years. Enforcing Fourier positivity is important for a consistent momentum-space interpretation of dipole amplitude and for its connection to the gluon content of the hadron~\cite{Gelis:2003vh,Blaizot:2004wu,Blaizot:2004wv,Dumitru:2005gt,Lappi:2013zma,Caucal:2025zkl}. }

{In fact, these tensions stem from the functional form assumed for the initial condition. Most analyses adopt a McLerran–Venugopalan (MV) form~\cite{McLerran:1993ni,McLerran:1993ka,McLerran:1994vd}, motivated by the large-$A$ limit of a nucleus, where many color sources per unit area ($\propto A^{1/3}$) yield Gaussian fluctuations and a Glauber–Mueller shape. For the \emph{proton} this motivation is weaker, and the MV parameters are typically fitted rather than fixed. Although improved BK evolution with higher-order impact factors has advanced the small-$x_B$ DIS description~\cite{Mantysaari:2018zdd,Hanninen:2022gje,Ducloue:2019jmy,Casuga:2025etc}, key tensions persist even in NLO Bayesian fits: the charm $\chi^2/\mathrm{d.o.f.}\gtrsim2.5$ and the negative regions in $S(k_T,x_B)$~\cite{Casuga:2025etc,Casuga:2023dcf}. It was also raised by Giraud and Peschanski~\cite{Giraud:2016lgg} as a possible deficiency of the dipole formalism. This suggests the initial-condition form, not the perturbative order alone, may be limiting. It motivates a more flexible yet dynamics-constrained dipole amplitude~\cite{Ducloue:2019jmy,Casuga:2025etc,Gao:2025dkn,Kou:2026iau}.  }

In this Letter we introduce a physics-informed neural network (PINN)~\cite{Raissi:2017zsi,9429985,DBLP:journals/corr/abs-2201-05624} to determine the universal dipole amplitude $N(r,x_B)$ as a differentiable function. Unlike a purely data-driven fit, the PINN enforces the small-$x$ evolution dynamics by including the collinearly improved BK (ciBK)~\cite{Beuf:2014uia,Iancu:2015vea,Iancu:2015joa,Ducloue:2019ezk,Ducloue:2019jmy} evolution residual in the loss function, while experimental data and basic physical constraints anchor the solution.  Rather than prescribing a parametric initial condition $N(r,x_0)$, we infer it nonparametrically from data under the evolution constraint. We perform a joint global analysis of DIS observables (including the reduced total, and reduced charm cross section) and exclusive diffractive $J/\psi$ photoproduction, with a  Fourier-positivity regularization imposed on the momentum-space dipole $S$-matrix. Uncertainties are quantified using deep ensembles~\cite{lakshminarayanan2017simplescalablepredictiveuncertainty}. We demonstrate that the extracted dipole amplitude simultaneously accommodates reduced total, charm quark cross sections,  exclusive vector meson constraints and remains consistent with Fourier-positivity in the data-constrained small-$x$ region.

\textbf{\textit{Physics-informed global analysis.}} 
In the small-$x$ limit, the dipole amplitude evolution is governed by the  BK equation. Beyond leading-logarithmic accuracy, large collinear corrections can destabilize the evolution~\cite{Iancu:2016pyc}. This instability is cured by the ciBK equation, which resums the dominant collinear enhancements to all orders~\cite{Beuf:2014uia,Iancu:2015vea,Iancu:2015joa,Ducloue:2019ezk,Ducloue:2019jmy}. In this work we adopt the ciBK prescription of Refs.~\cite{Ducloue:2019ezk,Ducloue:2019jmy,Iancu:2020jch}, formulated in the rapidity of the dense target and including the rapidity-shift (kinematic-constraint) structure of the evolution.
Unlike standard global fits that impose a specific analytic ansatz for the non-perturbative initial condition at a starting point $x_0$ and then evolve to smaller $x$, which can introduce parametrization bias, we represent the dipole amplitude across the full $(r,x_B)$ domain with a smooth surrogate, $N(r,Y)\equiv \mathcal{N}_\theta(r,Y),\qquad Y=\ln\!\left({x_0}/{x_B}\right)$.
Here $x_0$ is the starting point, $\theta$ denotes trainable parameters, and constrain $\mathcal{N}_\theta$ directly by the ciBK dynamics.
Concretely, we penalize the integro-differential residual
\begin{equation} \label{eq:ciBK} 
    \mathcal{R}_{\text{ciBK}}(r,Y)
    = \frac{\partial \mathcal{N}_{\theta}(r,Y)}{\partial Y}
    - \int d^2 \mathbf{r}_1\,
    K_{\text{ciBK}}(\mathbf{r},\mathbf{r}_1,\mathbf{r}_2)\,
    \mathcal{F}[\mathcal{N}_{\theta}],
\end{equation}
with $\mathbf{r}_2\equiv \mathbf{r}-\mathbf{r}_1$, and the residual $\mathcal{R}_{\text{ciBK}}(r,Y)$ enters the total objective in Eq.~\eqref{eq:loss_function} as a physics-based penalty that enforces the ciBK evolution for $\mathcal{N}_{\theta}(r,Y)$. Here $K_{\text{ciBK}}$ is the collinearly improved kernel and $\mathcal{F}$ denotes the nonlinear interaction functional. {Their explicit forms are given in the Supplemental Material. The amplitude $N(r,Y)$ in Eq.~\eqref{eq:ciBK} denotes the normalized impact-parameter-integrated dipole amplitude. We assume a factorized profile, so that $\sigma_{\rm dip}(r,Y)=2\int d^2b\,N(r,\mathbf{b},Y) =2\pi R_p^2 N(r,Y)=\sigma_0 N(r,Y)$. This factorized-profile approximation is justified in the dilute regime, and becomes approximate in the saturation region, where the saturation scale can develop a nontrivial impact-parameter dependence; see the recent discussion in Ref.~\cite{Li:2022avs,Kutak:2025wjy}.} 
{The ciBK equation employed here is built on the LO BK equation, supplemented by running-coupling corrections and the all-order resummation of the dominant logarithms~\cite{Beuf:2014uia,Iancu:2015vea,Iancu:2015joa,Ducloue:2019ezk,Ducloue:2019jmy}. While it does not include the full finite NLO BK kernel, this resummation serves as a practical prescription, given that the exact NLO BK equation is numerically unstable~\cite{Lappi:2016fmu}. By capturing the double- and single-logarithmic terms of the exact NLO kernel, this approach provides an approximation of the NLO+NLL evolution, establishing itself as an effective approach in small-$x$ phenomenology. A full NLO+NLL treatment, which also retains the genuine $\mathcal{O}(\alpha_s^2)$ non-logarithmic kernel terms, is prohibitively slow for a global PINN analysis and is left for future work. The running coupling uses the Brodsky–Lepage–Mackenzie prescription with a variable number of active flavors set by the dipole scale relative to the heavy-quark thresholds; see Sec.A1 of the Supplemental Material for details. } 
In evaluating  ciBK residual in Eq.~\eqref{eq:ciBK}, the rapidity shift makes the ciBK evolution non-local in $Y$, so one must specify a boundary condition for $N(r,Y<Y_0)$. In this work we adopt the frozen-boundary prescription, $N(r,Y<Y_0) = N(r, Y=Y_0)$.  We start the evolution at $x_0=0.03$, which reduces sensitivity to the boundary prescription and yields a smoother extrapolation towards $x \lesssim 10^{-2}$. The smooth neural surrogate also enables a direct evaluation of $\partial_Y N$ via automatic differentiation and stable Fourier transforms to momentum space, avoiding interpolation artifacts from a discrete $(r,Y)$ grid. 

The ciBK equation alone does not uniquely fix $\mathcal{N}_\theta$; we therefore determine it through a global fit to  the  DIS reduced total ($\sigma_r$),   charm quark ($\sigma_r^{c\bar c}$) cross section, and total exclusive  $J/\Psi$ photoproduction.  Even though our theoretical evolution initialized at higher momentum fractions, our actual fit to experimental data is strictly confined to the $x_B<0.004$ region. While $x_B \sim 0.01$ is widely regarded as the onset of gluon saturation, we restrict to $x_B<0.004$ to avoid the tail of the transitional region where DGLAP-dominated collinear and non-linear ciBK dynamics overlap~\cite{Albacete:2012rx,LHeC:2020van}. This conservative cut also mitigates the numerical impact of missing higher-order corrections in the DIS impact factors and reduces potential contamination from valence-quark contributions.
We employ an impact-parameter independent dipole amplitude and focus on total cross sections. For $J/\psi$ we include the standard real-part and skewness corrections \cite{Kovchegov:2012mbw,Golec-Biernat:1998zce,Kowalski:2003hm,Forshaw:2003ki,Kowalski:2006hc,Shuvaev:1999ce}, use the Boosted Gaussian wave function \cite{Kowalski:2006hc}, and allow an overall normalization $K_{\rm factor}^{\rm VM}$ \cite{Lappi:2013am,Mantysaari:2025ltq}. We take $m_{u,d,s}^{\rm eff}=0.14~\mathrm{GeV}$ and $m_c=1.4~\mathrm{GeV}$. 
Throughout, we employ leading-order impact factors, leaving a full NLO treatment for future work, since NLO calculations are several orders of magnitude slower than LO, particularly for heavy quarks computation, it involves up to a seven-dimensional numerical integral~\cite{Mantysaari:2018zdd,Hanninen:2022gje,Ducloue:2019jmy,Casuga:2025etc}. The  details of our calculations are presented in the  Supplemental Material.

\begin{figure*}[htbp] 
    \centering
        \centering
        \includegraphics[width=0.43\linewidth]{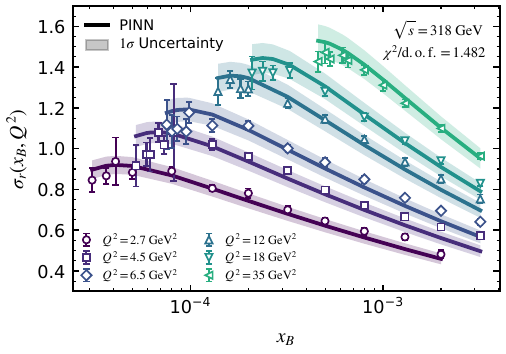}
        \includegraphics[width=0.43\linewidth]{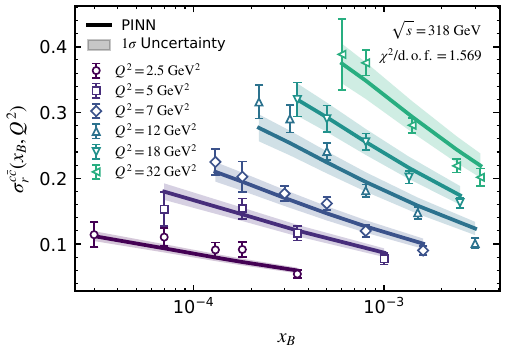}
    \caption{The reduced total and charm cross sections, $\sigma_r$ and $\sigma_r^{c\bar c}$, are shown vs. $x_B$ in selected $Q^2$ bins at $\sqrt{s}=318~\mathrm{GeV}$, compared with HERA data \cite{H1:2015ubc,H1:2009pze,H1:2018flt}. Solid curves are the best-fit PINN prediction (minimum global $\chi^2/\mathrm{d.o.f.}$); bands show $1\sigma$ uncertainties. The quoted $\chi^2/\mathrm{d.o.f.}$ uses the solid curve; for $\sigma_r$ it includes all available energies.}  
    \label{fig:sigma_ep}
\end{figure*}

\begin{figure}[htbp]
    \centering
    \includegraphics[width=0.43\textwidth]{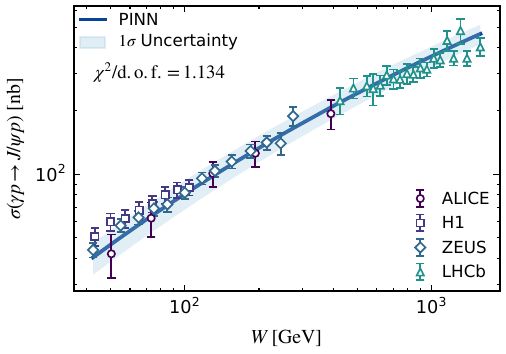}
    \caption{Total cross section for exclusive $J/\psi$ photoproduction as a function of center-of-mass
energy  $W$, compare to ALICE~\cite{ALICE:2014eof,ALICE:2018oyo}, H1~\cite{H1:2005dtp,H1:2013okq}, ZEUS~\cite{ZEUS:2002wfj} and LHCb~\cite{LHCb:2018rcm,LHCb:2024pcz} data. Solid curve is the PINN prediction with the smallest global $\chi^2/\mathrm{d.o.f.}$; shaded band is $1\sigma$ uncertainties. The quoted $\chi^2/\mathrm{d.o.f.}$ value is evaluated using the solid curve.}
    \label{fig:Jpsi}
\end{figure}

\begin{figure}[htbp]
    \centering
    \includegraphics[width=0.43\textwidth]{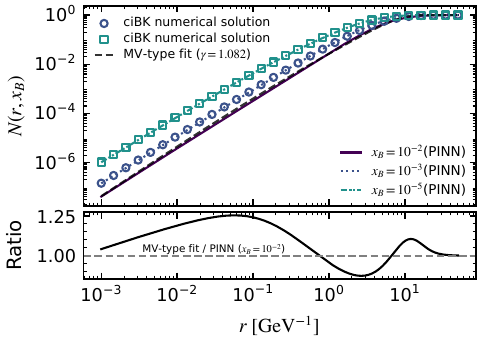}
    \caption{Top: Solid curves show the dipole  amplitude $N(r,x_B)$ predicted by the PINN for $x_B \in \{10^{-2},\,10^{-3},\,10^{-5}\}$. Markers denote numerical ciBK evolution results obtained with the same initial condition as used in the PINN. The dashed curve corresponds to the best-fit MV-type parametrization to $N_{\rm PINN}(r,x_B=0.01)$.}
    \label{fig:BK_solution}
\end{figure}

\begin{figure}[htbp]
    \centering
    \includegraphics[width=0.43\textwidth]{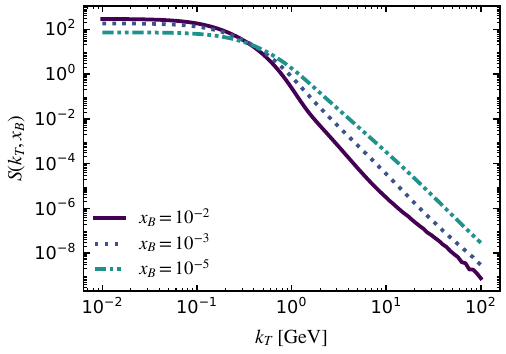}
    \caption{Fourier transform of the dipole amplitude at $x_B \in \{10^{-2},\ 10^{-3},\ 10^{-5}\}$.}
    \label{fig:FT}
\end{figure}

We further impose essential physical constraints through penalty terms. In the large-dipole limit ($r\to r_{\max}$), we implement saturation unitarity requirement by enforcing the black-disk limit $N(r_{\max},Y)\to 1$~\cite{Albacete:2004gw,Marquet:2005ic}. {In the small-dipole limit ($r\to r_{\min}$), color transparency requires $N(r,Y)\propto r^{2\gamma}$~\cite{Munier:2003vc} with $\gamma$  the  anomalous dimension.  Based on the Bochner theorem~\cite{bochner}, Fourier positivity of the dipole amplitude implies the necessary condition $0 < \gamma \le 1$~\cite{Giraud:2016lgg}.} Rather than fixing a power-law ansatz for $N(r,Y)$, we impose this  condition on the local logarithmic slope, $\gamma = \partial \ln N(r,Y)/\partial \ln r^2$ by requiring $0 < \gamma \le 1$ in the small-dipole region $r<10^{-3}\, \mathrm{GeV}^{-1}$ for $Y\ge\ln(0.03/0.01)\simeq 1.1$. Since this anomalous-dimension bound is necessary but not sufficient for Fourier positivity, we also directly penalize negative values of the momentum-space dipole $S$-matrix,
\begin{align}\label{eq:FT}
S(k_T,Y)=\int d^2\mathbf re^{i\mathbf k_T\cdot\mathbf r}\,\bigl[1-N(r,Y)\bigr].
\end{align}

The total objective in our PINN now reads
\begin{equation}\label{eq:loss_function}\mathcal{L}(\theta,\mathbf{p})=w_1\mathcal{L}_{\text{ciBK}}+w_2\mathcal{L}_{\text{data}}+w_3\mathcal{L}_{\text{phy}},
\end{equation}
{where \(w_i\) are numerical balancing factors controlling the relative importance of each term. Their values are determined through gradient-based loss-weight calibration, as described in Sec.~A3 of the Supplemental Material.} The
$\mathcal{L}_{\text{ciBK}}=\|\mathcal{R}_{\text{ciBK}}\|^2$ is the ciBK residual defined in Eq.~\eqref{eq:ciBK}, $\mathcal{L}_{\text{data}}$ quantifies agreement with the experimental data, and $\mathcal{L}_{\text{phy}}$ encodes additional physical constraints.  
Optimization is performed with the AdamW algorithm~\cite{loshchilov2018decoupled} together with a cosine-annealing learning-rate schedule~\cite{DBLP:journals/corr/LoshchilovH16a},  ensuring stable and high-precision convergence. {We simultaneously fit the network parameters $\theta$ and the physical parameters $\mathbf{p}=\{R_p,\,K_{\rm factor}^{VM},\,C^2\}$. Here $R_p$ is the transverse proton size, entering the normalization through $\sigma_0/2=\pi R_p^2$ with $\sigma_{\rm dip}(r,x_B)=\sigma_0\, N(r,x_B)$; $K_{\rm factor}^{VM}$ multiplies the exclusive vector-meson cross section to absorb residual uncertainties in the non-perturbative $J/\psi$ wave function; and $C^2$ fixes the running-coupling scale $\mu^2=4C^2/r^2$ in the ciBK kernel. The infrared-frozen coupling is fixed to $\alpha_{\rm fr}=0.7$.}
In this framework, the effective anomalous dimension is determined locally from gradients of $\mathcal{N}_\theta(r,Y)$, rather than treated as a fixed constant. 
Uncertainties are quantified with a hybrid strategy combining the Monte Carlo replica method and a deep ensemble approach~\cite{lakshminarayanan2017simplescalablepredictiveuncertainty}. To propagate experimental errors, we generate statistically independent pseudodata replicas by sampling each data point from a Gaussian distribution centered on the measured central value with a width given by its uncertainty, and train an independent network for each replica. Each network outputs both the continuous amplitude $N(r,Y)$ and a data-dependent predictive variance needed to construct theory predictions. We therefore take $\mathcal{L}_{\text{data}}$ to be a negative log-likelihood, which naturally incorporates heteroscedastic (aleatoric) uncertainties. This yields a robust uncertainty band for the extracted universal amplitude $N(r,Y)$. Detailed expressions for $\mathcal{L}_{\text{data}}$ and $\mathcal{L}_{\text{phy}}$, together with a full description of the PINN framework, are provided in the Supplemental Material.

\textbf{\textit{Results and Discussion.}} 
The global fit includes the HERA reduced cross section $\sigma_r$ at all available energies $\sqrt{s}=224.9,\,251.5,\,300.3,$ and $318~\mathrm{GeV}$~\cite{H1:2015ubc,H1:2009pze}, the reduced charm cross section $\sigma_r^{c\bar c}$ at $\sqrt{s}=318~\mathrm{GeV}$~\cite{H1:2018flt}, and exclusive diffractive $J/\psi$ photoproduction~\cite{ALICE:2014eof,ALICE:2018oyo,H1:2005dtp,H1:2013okq,ZEUS:2002wfj,LHCb:2018rcm,LHCb:2024pcz}. Unless stated otherwise, we restrict to the measurements with $x_B\le 0.004$ and $2<Q^2<45~\mathrm{GeV}^2$.

Figure~\ref{fig:sigma_ep} shows representative results at $\sqrt{s}=318~\mathrm{GeV}$ for  $\sigma_r$ and $\sigma_r^{c\bar c}$. {The predictions obtained from the best PINN-extracted amplitude describe the reduced total cross-section data with $\chi^2/N_{\rm dof}=1.482$, while maintaining a good description of the charm reduced cross section with $\chi^2/N_{\rm dof}=1.569$.} The same joint optimization simultaneously determines the phenomenological parameters (deep-ensemble $1\sigma$) as {$R_p=0.870\pm 0.048~\mathrm{fm}$, $K_{\mathrm{factor}}^{VM}=1.079\pm 0.020$ and $C^2=0.955\pm 0.225$}. Reconciling DIS reduced total  and charm observables within small-$x$ evolution frameworks is known to be challenging. Recent analyses~\cite{Hanninen:2022gje,Casuga:2025etc} emphasize that charm production probes short dipole distances and therefore provides particularly stringent constraints on the short-distance behavior and on the  profile entering the evolution. Our simultaneous fit yields a universal dipole amplitude together with physically reasonable parameter values. Further detailed fit results for $\sigma_r$ at various center-of-mass energies, as well as the evaluation of the longitudinal structure function $F_L$ serving as a crucial cross-check, are provided in the Supplemental Material. 

The extracted amplitude also describes diffractive $J/\psi$ production within the same global framework. Figure~\ref{fig:Jpsi} shows the total cross section for $\gamma p\to J/\psi\,p$ as a function of the photon--proton energy $W$, compared with experimental data~\cite{ALICE:2014eof,ALICE:2018oyo,H1:2005dtp,H1:2013okq,ZEUS:2002wfj,LHCb:2018rcm,LHCb:2024pcz}. {The PINN prediction provides an excellent description, with $\chi^2/N_{\rm dof}=1.134$. The overall normalization is fixed to $K_{\mathrm{factor}}^{VM}=1.079$ to account for uncertainties in the vector-meson wave function}, and no additional process-dependent initial conditions are introduced. This offers a nontrivial universality check of the inferred $N(r,x_B)$ across inclusive DIS (including charm) and exclusive vector-meson photoproduction.

Figure~\ref{fig:BK_solution} shows the extracted dipole amplitude $N(r,x_B)$ at three representative $x_B$ values. As a consistency check, we solve the ciBK equation numerically using $N(r,x_B=x_0)=N_{\rm PINN}(r,x_B=x_0)$ with $x_0=0.03$ as the initial condition to consistently accommodate the non-local rapidity shift in the evolution kernel, and compare the evolved solutions with PINN output. We observe excellent agreement throughout the kinematic regions. Once trained, the surrogate evaluates $N(r,x_B)$ at any desired $x_B$ in a single forward pass, avoiding repeated and computationally expensive grid-based ciBK evolutions in global analyses. The inferred pre-asymptotic shape, together with the momentum-space positivity regularization, is further illustrated by comparing $N_{\rm PINN}(r,x_B=0.01)$, a standard reference scale in saturation phenomenology, with an MV-type parametrization~\cite{McLerran:1993ni,McLerran:1993ka},
\begin{equation}\label{eq:MV_model}
N_{\text{MV-type}}(r)
= 1 - \exp\left[
-\frac{(r^2 Q_{s0}^2)^\gamma}{4}
\ln\left(\frac{1}{r\Lambda} + e_c \cdot e\right)
\right].
\end{equation}
{Fitting Eq.~\eqref{eq:MV_model} to the PINN profile at $x_B = 0.01$ yields
$(\gamma,\ Q_{s0}^2,\ \Lambda,\ e_c)=(1.082,\ 0.080~\mathrm{GeV}^2,\ 0.283~\mathrm{GeV},\ 0.649)$.
The ratio of the fitted MV-type form to the PINN result, shown in the bottom panel of Figure~\ref{fig:BK_solution}, exhibits systematic deviations of up to $\sim 25\%$ over the range $r \in [10^{-2},\ 10^{-1}]~\mathrm{GeV}^{-1}$. This indicates that the extracted initial dipole amplitude cannot be faithfully captured by the MV-type functional form.} By moving beyond rigid parametrizations, the non-parametric PINN extraction provides the flexibility needed to capture the dipole-amplitude structure at both short and long transverse distances, which is essential for a simultaneous description of light- and heavy-quark observables and the momentum-space positivity requirement.

Finally, we examine the two-dimensional Fourier transform of the dipole $S$-matrix defined in Eq.~\eqref{eq:FT}. Momentum-space positivity is essential: within the CGC framework, the momentum-space dipole is directly connected to the gluon content of the target and enters a variety of particle-production cross sections in $e$--$p/A$ and $p$--$p/A$ collisions. The sign changes in the momentum-space dipole have been reported for several widely used initial conditions, even when the DIS reduced cross section is well described~\cite{Albacete:2010sy,Lappi:2013zma,Ducloue:2019jmy,Casuga:2025etc,Caucal:2025zkl}. Although BK evolution generally mitigates these issues, they can persist in phenomenologically relevant kinematics. Figure~\ref{fig:FT} shows $S(k_T,x_B)$ at three representative Bjorken-$x_B$ values. In our PINN global fit we impose a positivity regularization by penalizing regions where $S(k_T,x_B)<0$. In the region ($x_B\le 0.01$), the extracted PINN solution yields a smooth $k_T$ dependence and remains positive over the range shown, $k_T\in[0,100]~\mathrm{GeV}$. This demonstrates that momentum-space positivity can be achieved simultaneously with a good description of inclusive DIS reduced cross section (including total and charm quark) and diffractive vector-meson production, provided sufficient functional flexibility and an explicit positivity regularization are included. {We have also checked that relaxing the $S(k_T)>0$ constraint does yield a smaller $\chi^2/{\rm d.o.f}$, but at the price of a positivity-violating $S$-matrix. The constraint therefore refines the extraction, yielding a dipole amplitude that is at once data-driven and physically admissible.} For comparison, Ref.~\cite{Casuga:2025etc} reports oscillatory behavior in $S(k_T,x_B)$ even at $x_B=10^{-3}$ and $10^{-6}$ using a Bayesian-fit approach.

\textbf{\textit{Conclusion.}} We have performed a global and dynamics-constrained determination of the universal small-$x$ dipole amplitude $N(r,x_B)$ using a physics-informed neural network. In our approach, the amplitude is represented by a smooth surrogate $\mathcal{N}_\theta(r,x_B)$ and is constrained directly at the functional level by the ciBK evolution equation together with precision DIS and diffractive data. The initial profile is inferred non-parametrically from data while remaining consistent with ciBK dynamics. 
{Within the fitted kinematic region, the resulting universal amplitude simultaneously describes the reduced total cross section, the reduced charm cross section, and exclusive $J/\psi$ photoproduction. Another central outcome is a well-behaved momentum-space representation: by incorporating the Fourier-positivity penalty on the momentum-space $S$-matrix, we obtain a PINN solution whose $S(k_T,x_B)$ remains non-negative for $x_B \le 0.01$ over the transverse-momentum range shown. This addresses a long-standing issue in dipole-model initial conditions and enables consistent use of the extracted amplitude in downstream CGC calculations requiring a positive momentum-space input. It also motivates going beyond the McLerran–Venugopalan parametrization as an initial condition for the proton.}

The present study employs an impact-parameter independent dipole amplitude and leading-order impact factors, and we restrict the fit to $x_B\le 0.004$ experimental data to suppress potential DGLAP-dominated contributions and reduce sensitivity to pre-asymptotic effects associated with the rapidity-shift non-locality. {Natural next steps are to add impact-parameter dependence and NLO impact factors with NLO + NLL evolution; beyond the mean-field BK level, the LO~\cite{Mueller:2001uk,Lappi:2012vw,Schlichting:2014ipa} and NLO JIMWLK~\cite{Kovner:2013ona, Kovner:2014lca} equations can also be solved with the score-based method and linked to our PINN framework. Together, these would enable more stringent universality tests and direct confrontation with forthcoming EIC precision data.}

More broadly, our framework moves beyond a purely ``data-driven” fit toward ``physics-driven” inference. PINN acts as a differentiable surrogate that simultaneously enforces the evolution equation, accommodates precise measurements, and incorporates first-principles constraints such as unitarity and color transparency. By combining these complementary sources of information, the network is guided not only by statistical agreement with data but also by the underlying QCD dynamics, helping ensure that the extracted amplitude remains physically consistent across the full kinematic domain. This strategy is inherently generalizable and offers a robust template for embedding fundamental laws into data-driven inference in scientific settings where dynamical equations and first-principles constraints coexist with experimental measurements.

We have made available the tabulated PINN solution for $N(r,x_B)$ over the fitted domain, together with its momentum-space Fourier transform, in Ref.~\cite{dai_2026_20794812}.

\begin{acknowledgments}

\textbf{\textit{Acknowledgments}} We thank {Haowu Duan}, Heikki M\"antysaari, {Farid Salazar},  and Dionysis N. Triantafyllopoulos  for helpful comments. This work was supported in part by the NSFC under grant No.~12225503, No.~12535010, No.~12435009, No.~12405156, and by the Outstanding Leading Talent Team Program of Central China Normal University (XJ2026000302). 
F.P. L acknowledges support from the NSFC under grant No.~12325507, No.~12547102, and No.~12147101, and the National Key Research and Development Program of China under grant No. 2022YFA1604900. This work was supported also in part by the Cross Research Project of Fundamental Research Funds for Central Universities of Central China Normal University in 2025: ``Advanced Detection and Artificial Intelligence at the Frontiers of Physics'' (No.30101250317).

\end{acknowledgments}
\bibliographystyle{apsrev}
\bibliography{refs}

\clearpage
\section{SUPPLEMENTAL MATERIAL}
\setcounter{equation}{0}
\renewcommand{\theequation}{S\arabic{equation}}
\setcounter{figure}{0}
\renewcommand{\thefigure}{S\arabic{figure}}

\begin{figure*}
    \centering
    \includegraphics[width=0.9\textwidth]{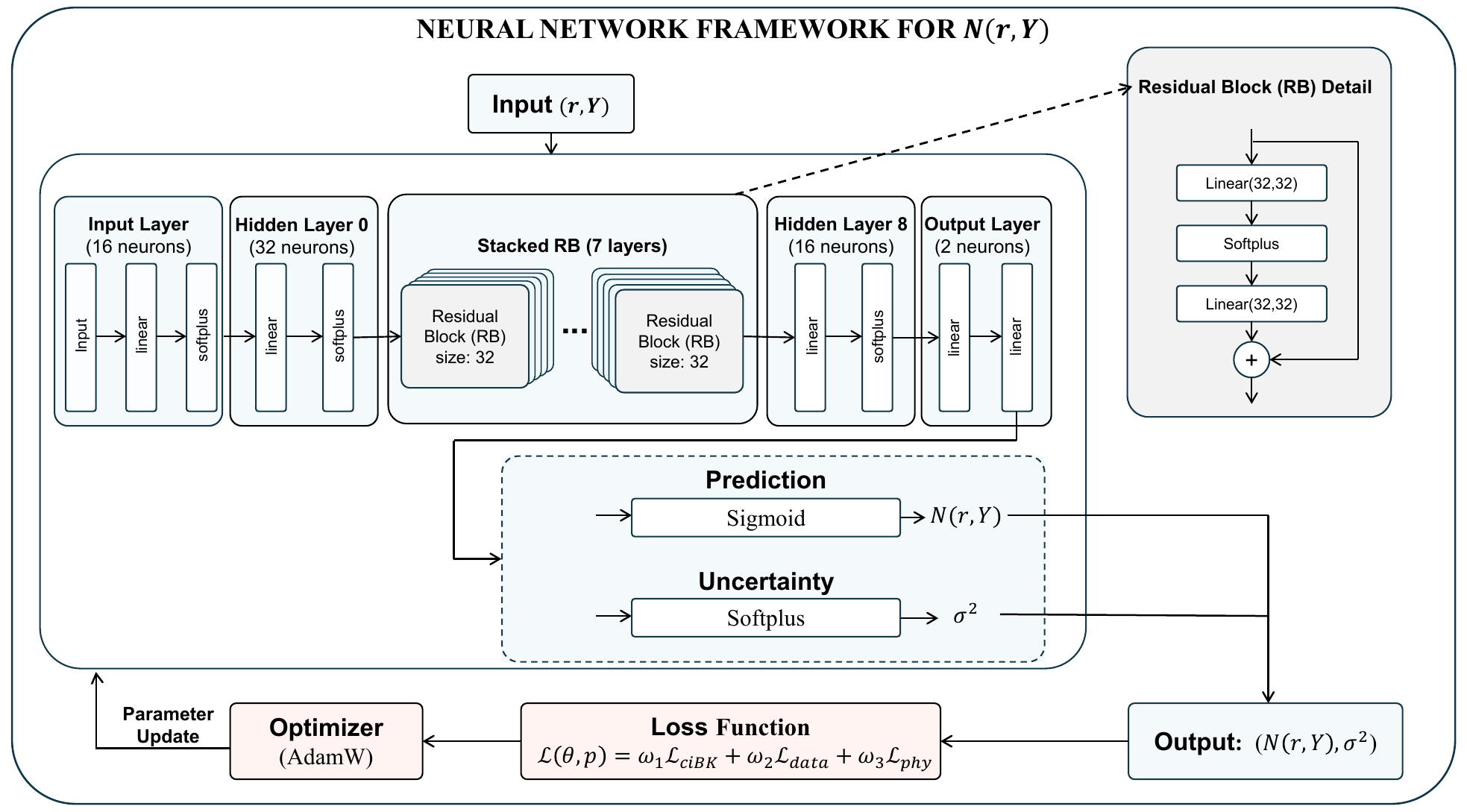}
    \caption{Schematic of the PINN framework. The network maps the kinematic inputs $(r,Y)$ to the dipole amplitude $N(r,Y)$. Training is guided by the total loss $\mathcal{L}(\theta,\mathbf{p})=w_1\mathcal{L}_{\text{ciBK}}+w_2\mathcal{L}_{\text{data}}+w_3\mathcal{L}_{\text{phy}}.$ See the text for details of the three loss components.}
    \label{fig:PINN}
\end{figure*}

\subsection{Physics-Informed Framework for Dipole Evolution}
This Supplementary Material section provides the details about the PINN framework used to extract the universal dipole scattering amplitude. We first outline the { collinearly-improved Balitsky-Kovchegov (ciBK) equation that governs the small-$x$ evolution and the running coupling}, and then detail the neural network architecture along with the formulation of the loss function.

\subsubsection*{A1.The Collinearly-Improved Balitsky--Kovchegov Equation and the Running Coupling}
{{The small-$x$ evolution of the dipole scattering amplitude $N(r,Y)$ is governed by the Balitsky--Kovchegov (BK) equation. To achieve high-precision phenomenology compatible with experimental data, we employ the collinearly-improved BK (ciBK) equation, which resums large collinear logarithms to all orders and enforces exact kinematic constraints. The evolution equation in rapidity $Y$ reads~\cite{Ducloue:2019ezk, Ducloue:2019jmy}
\begin{align} \label{eq:ciBK_form}
\frac{\partial N(r,Y)}{\partial Y} ={}& \frac{\bar{\alpha}_{\text{BLM}}(r)}{2 \pi} \int d^2 \mathbf{r}_1 \frac{r^2}{r_1^2 r_2^2} \left[ \frac{r^2}{\min(r_1^2, r_2^2)} \right]^{\pm A_1 \bar{\alpha}_{\text{BLM}}}  \notag \\
&\times \mathcal{K}_{\text{DLA}}(\rho) \Big[ N(r_1, Y-\delta_1) + N(r_2, Y-\delta_2)  \notag \\
&-N(r, Y) - N(r_1, Y-\delta_1)N(r_2, Y-\delta_2) \Big],
\end{align}
where $\mathbf{r}_2 = \mathbf{r} - \mathbf{r}_1$ and the evolution kernel factorizes into three components: the leading-order dipole kernel supplemented by the running-coupling prescription, the DLA factor resumming the double collinear logarithms to all orders, and the single transverse logarithmic (STL) correction factor. Here $\bar{\alpha}_{\text{BLM}}$ denotes the Brodsky--Lepage--Mackenzie (BLM) running coupling~\cite{Brodsky:1982gc}, whose explicit three-dipole form is given in Eq.~\eqref{eq:alpha_BLM}. These four elements are defined as follows.

\textbf{(i) The kinematic constraints (rapidity shifts):}

The lifetime ordering of successive soft-gluon emissions, required for the consistency of the evolution at small $x$, is enforced through the rapidity shifts entering the arguments of the amplitudes in the nonlinear functional,
\begin{equation}
    \delta_i = \max \left\{0, \ln \frac{r^2}{r_i^2} \right\}, \qquad i=1,2,
\end{equation}
where $r_1$ and $r_2$ denote the sizes of the two daughter dipoles produced in the emission and $r$ is the size of the parent dipole. These constraints relate the rapidity available to the daughter dipoles to their transverse sizes: an emission into a much smaller daughter dipole incurs a finite rapidity shift, suppressing configurations that would otherwise violate time ordering. In this way the shifts resum the dominant double collinear logarithms generated by strictly time-ordered emissions to all orders, while leaving the conventional BK kernel structure unchanged.

\textbf{(ii) The double logarithmic approximation (DLA) factor.}

In a formulation without the kinematic constraints, the same double collinear logarithms are resummed in closed form by the DLA factor, given by the Bessel function
\begin{equation}
\mathcal{K}_{\text{DLA}}(\rho) = \frac{J_1(2\sqrt{\bar{\alpha}_s \rho^2})}{\sqrt{\bar{\alpha}_s \rho^2}}  = 1 - \frac{\bar{\alpha}_s \rho^2}{2} + \frac{(\bar{\alpha}_s \rho^2)^2}{12} + \cdots,
\end{equation}
where $\rho = \sqrt{L_{r_1 r}\, L_{r_2 r}}$ with $L_{r_i r} = \ln(r_i^2 / r^2)$. Expanding the shifted amplitudes $N(r_i, Y-\delta_i)$ in powers of $\bar{\alpha}_s$ reproduces, order by order, the same double-logarithmic tower $(\bar{\alpha}_s \rho^2)^n$ that is resummed by $\mathcal{K}_{\text{DLA}}(\rho)$. The rapidity shifts $\delta_i$ and the DLA factor therefore constitute two equivalent prescriptions for resumming the same double collinear logarithms, and retaining both simultaneously would double-count these contributions. Since the double-logarithmic resummation is already fully carried by the kinematic constraints $\delta_i$, we keep only the leading term, $\mathcal{K}_{\text{DLA}}(\rho) = 1$. This choice stabilizes the evolution speed and avoids the spurious oscillations associated with the higher-order terms of the Bessel expansion.

\textbf{(iii) The single transverse logarithmic (STL) correction factor.}

Beyond the double collinear logarithms, the next-to-leading-order corrections to the BK kernel also generate single transverse logarithms, $\ln(r^2/r_i^2)$, which are not captured by the time-ordering constraints. These are resummed to all orders by the STL correction factor,
\begin{equation}
\mathcal{K}_{\text{STL}} = \left[ \frac{r^2}{\min(r_1^2, r_2^2)} \right]^{\pm A_1 \bar{\alpha}_{\text{BLM}}}, \qquad A_1 = \frac{11}{12},
\end{equation}
where the coefficient $A_1$ originates from the one-loop DGLAP anomalous dimension. The sign in the exponent is fixed by the relative size of the dipoles: the plus sign applies when $r^2 < \min(r_1^2, r_2^2)$ and the minus sign when $r^2 > \min(r_1^2, r_2^2)$, so that the factor consistently suppresses the kernel in the region of strongly ordered dipole sizes. Together with the kinematic constraints and the leading DLA term, this factor completes the collinear improvement of the BK kernel adopted in the present analysis.

\textbf{(iv) The running coupling prescription.}

The running-coupling corrections form an important and well-understood class of the NLO contributions to the BK equation, being the subset enhanced by the one-loop $\beta$-function. Among the prescriptions proposed in the literature, we adopt the BLM prescription (also dubbed as ``fast apparent convergence"~\cite{Iancu:2015joa}). In this prescription, the effective coupling entering the kernel is written as a smooth interpolating coupling $\bar{\alpha}_{\text{BLM}}(r,r_1,r_2)$, which depends on the sizes of all three dipoles involved in the splitting. The detailed BLM formula is the Eq.~(\ref{eq:alpha_BLM}). This prescription absorbs the dominant $\beta_0$-dependent part of the NLO corrections into the kernel and effectively selects the smallest dipole as the relevant scale, thereby minimizing the residual higher-order corrections. It is defined as
\begin{equation} \label{eq:alpha_BLM}
\bar{\alpha}_{\text{BLM}}(r, r_1, r_2) = \left[
\frac{1}{\bar{\alpha}_s(r)} + \frac{r_1^2 - r_2^2}{r^2} \frac{\bar{\alpha}_s(r_1) - \bar{\alpha}_s(r_2)}{\bar{\alpha}_s(r_1) \bar{\alpha}_s(r_2)} \right]^{-1},
\end{equation}
with $\bar{\alpha}_s(r) \equiv N_c\,\alpha_s(r)/\pi$. In any limit where one of the three dipoles is much smaller than the other two, Eq.\eqref{eq:alpha_BLM} reduces to the minimal-dipole choice $\bar{\alpha}_s(r_{\min})$.

The one-loop coupling $\alpha_s(r)$ is evaluated at the coordinate-space scale $\mu^2 = 4C^2/r^2$, where the dimensionless constant $C$ originates from the Fourier transform between coordinate and momentum space and is treated as a free parameter in the global fit. The coupling incorporates heavy-quark flavor thresholds: the number of active flavors increases from $n_f=3$ to $4$ and to $5$ as $\mu^2$ crosses $m_c^2$ and $m_b^2$, respectively. To regularize the Landau pole in the infrared, we freeze the coupling through a smooth shift $\lambda_0$~\cite{Albacete:2004gw},
\begin{equation}
\alpha_s(r) = \frac{1}{\beta_0^{(n_f)} \ln\left( \frac{4C^2}{r^2 \Lambda_{(n_f)}^2} + \lambda_0 \right)}, \quad \beta_0^{(n_f)} = \frac{11N_c - 2n_f}{12\pi}.
\end{equation}
with $\lambda_0 = \exp\!\big[1/(\beta_0^{(3)}\times\alpha_{\text{fr}})\big]$, which ensures that the coupling saturates smoothly to a maximal value $\alpha_{\text{fr}}$ in the deep infrared. Continuity of $\alpha_s(r)$ across the thresholds $\mu^2 = m_f^2$ ($f=c,b$) is enforced by recursively matching the scale parameters,
\begin{equation}
\Lambda_{(n_f-1)} = m_f \left[ \left( \frac{m_f^2}{\Lambda_{(n_f)}^2} + \lambda_0 \right)^{\beta_0^{(n_f)} \over \beta_0^{(n_f-1)}} - \lambda_0 \right]^{-1/2}.
\end{equation}
The matching is anchored at the $Z$ pole, where $\Lambda_{(5)}$ is fixed by the world-average $\alpha_s(M_Z)$. Throughout the global extraction we fix $N_c = 3$, $m_c = 1.4~\text{GeV}$, $m_b = 4.18~\text{GeV}$, $M_Z = 91.2~\text{GeV}$, $\alpha_s(M_Z^2) = 0.118$~\cite{ParticleDataGroup:2026aaa}, and the frozen coupling $\alpha_{\text{fr}} = 0.7$~\cite{Albacete:2010sy}.

}}

\subsubsection*{A2. Physics-Informed Neural Network Framework and Loss Function}

Figure~\ref{fig:PINN} illustrates the PINN architecture, which takes the dipole size $r$ and the rapidity $Y=\ln(x_0/x_B)$ as inputs. The network is based on a Residual Neural Network (ResNet)~\cite{DBLP:journals/corr/HeZRS15}, which mitigates vanishing-gradient issues in deep multilayer perceptrons. Our baseline configuration comprises seven residual blocks with 32 neurons each, using the softplus activation $\ln(1+e^x)$ for the hidden nonlinearities. A final sigmoid activation, $\sigma(x)=1/(1+e^{-x})$, constrains the output to $(0,1)$, ensuring that the dipole amplitude $N(r,Y)$ remains physical.

The optimization objective, i.e., the loss function, is formulated as:
\begin{equation}\label{eq:loss}\mathcal{L}(\theta,\mathbf{p})=w_1\mathcal{L}_{\text{ciBK}}+w_2\mathcal{L}_{\text{data}}+w_3\mathcal{L}_{\text{phy}},
\end{equation}
where $\theta$ denotes the trainable neural network weights and biases, and $\mathbf{p}$ represents optimizable physical parameters. The weights $w_i$ are numerical balancing factors controlling the relative importance of the different loss terms. They are determined through gradient-level calibration before production training and then held unchanged during the subsequent optimization.

The term $\mathcal{L}_{\text{ciBK}}=\|\mathcal{R}_{\text{ciBK}}\|^2$ is the ciBK residual, which is defined as
\begin{equation} 
    \mathcal{R}_{\text{ciBK}}(r,Y)
    = \frac{\partial \mathcal{N}_{\theta}(r,Y)}{\partial Y}
    - \int d^2 \mathbf{r}_1\,
    K_{\text{ciBK}}(\mathbf{r},\mathbf{r}_1,\mathbf{r}_2)\,
    \mathcal{F}[\mathcal{N}_{\theta}].
\end{equation}

Here, the collinear-improved kernel incorporates the BLM prescription along with the resummation factors discussed above:
\begin{equation}
    K_{\rm ciBK}(\mathbf{r},\mathbf{r}_1,\mathbf{r}_2) = \frac{\bar{\alpha}_{\text{BLM}}(r)}{2\pi}\,\frac{r^2}{r_1^2 r_2^2}\,\mathcal{K}_{\rm DLA}(\rho)\,\mathcal{K}_{\rm STL}.
\end{equation}

The associated nonlinear functional $\mathcal{F}[\mathcal{N}_\theta]$ enforces the exact kinematic constraints:
\begin{align}
    \mathcal{F}[\mathcal{N}_\theta] ={}& 
    \mathcal{N}_\theta(r_1, Y-\delta_1) + \mathcal{N}_\theta(r_2, Y-\delta_2) - \mathcal{N}_\theta(r, Y)\notag \\ 
    &- \mathcal{N}_\theta(r_1, Y-\delta_1)\mathcal{N}_\theta(r_2, Y-\delta_2).
\end{align}

During the network optimization, this continuous residual is evaluated on a set of collocation points generated via two-dimensional Latin hypercube sampling (LHS) over the $(r,Y)$ domain. These points are resampled at the start of each training epoch. Compared with a dense, fixed uniform grid, this dynamic resampling reduces the per-epoch computational cost while ensuring thorough coverage of the continuous kinematic space over the course of training.

The data loss $\mathcal{L}_{\text{data}}$ measures agreement with experimental data. Consistent with the deep-ensemble setup, we use a negative log-likelihood,
\begin{align} \label{eq:loss_data}
   \mathcal{L}_{\text{data}} =& \sum_{\mathcal{O}}\sum_{i=1}^{N_{\mathcal{O}}} \lambda_{\mathcal{O}} \Bigg[ \frac{(\mathcal{O}_i^{\text{th}} - \mathcal{O}_i^{\text{exp}})^2}{2\big((\sigma_i^{\text{NN}})^2 + (\sigma_i^{\text{exp}})^2\big)} \notag \\ 
    &+ \frac{1}{2}\ln\big( (\sigma_i^{\text{NN}})^2+(\sigma_i^{\text{exp}})^2 \big) \Bigg]\,,
\end{align}
where the outer sum runs over observable classes $\mathcal{O}\in\{\sigma_r,\sigma_r^{c\bar c},\sigma^{\text{VM}}\}$, and $\lambda_{\mathcal{O}}$ are dataset-specific weights. These weights are introduced to balance the relative contributions of different observables, preventing the overall data loss from being strictly dominated by datasets with a significantly larger number of data points (e.g., the inclusive $\sigma_r$). For the $i$-th point, $\mathcal{O}_i^{\text{exp}}$ and $\sigma_i^{\text{exp}}$ are the measured central value and total experimental uncertainty. The corresponding prediction $\mathcal{O}_i^{\text{th}}$ is computed from the network output, namely the universal dipole amplitude $N(r,Y)$. The term $\sigma_i^{\text{NN}}$ is the data-dependent predictive uncertainty produced by the network’s variance branch during training.

The physical loss $\mathcal{L}_{\text{phy}}$ imposes key theoretical constraints, including the large-$r$ limit, momentum-space positivity, and the small-$r$ color-transparency behavior. Its explicit form is
\begin{align} \label{eq:loss_phy}
    \mathcal{L}_{\text{phy}} &= \lambda_{\text{IR}}\| N(r_{\text{max}},Y) - 1 \|_{r_{\text{max}}= 50\, \mathrm{GeV}^{-1}}^2  \notag \\
    &\quad + \lambda_{\text{pos}} \| \min(0, S(k_T,Y)) \|^2 \notag \\
    &\quad + \lambda_{\text{UV}} \Bigg( \left\| \max\left(0, \frac{\partial \ln N(r,Y)}{\partial \ln r^2} - 1\right) \right\|^2 \notag \\ 
    &\quad + \left\| \min\left(0, \frac{\partial \ln N(r,Y)}{\partial \ln r^2}\right) \right\|^2 \Bigg)_{r<10^{-3}\,\mathrm{GeV}^{-1}}\, ,
\end{align}
where the penalty terms enforce the following physical behaviors:
\begin{itemize}
    \item \textbf{Asymptotic large-$r$ constraint (Black-disk limit):} Because $r_{\text{max}}=50~\mathrm{GeV}^{-1}$ (about $10~\mathrm{fm}$) is far larger than the proton radius, the dipole amplitude should saturate to the black-disk limit. We therefore penalize deviations from $N(r_{\text{max}},Y)=1$.
    \item \textbf{Asymptotic small-$r$ constraint (Color transparency):} In the deep perturbative regime ($r<10^{-3}~\mathrm{GeV}^{-1}$), color transparency implies $N(r,Y)\sim r^{2\gamma}$, where $\gamma$ is an effective anomalous dimension. The third term constrains the local logarithmic slope $\gamma \equiv \partial \ln N(r,Y)/\partial \ln r^2$ to the physically allowed range $[0,1]$, preserving the small-$r$ asymptotics while retaining the functional flexibility required by QCD evolution.
    \item \textbf{Fourier-positivity constraint:} The second term penalizes negative values of the momentum-space $S$-matrix $S(k_T,Y)$, enforcing positivity of the Fourier-transformed distribution during training.
\end{itemize}

{Here, $\lambda_{\text{IR}}$, $\lambda_{\text{pos}}$, and $\lambda_{\text{UV}}$ are penalty coefficients that control the relative strength of the corresponding physical constraints. These coefficients, together with $w_i$, are numerical balancing factors rather than additional physical fit parameters. In the default production training, the loss weights are kept fixed and set to $(w_1,\ w_2,\ w_3) = (10000,\ 1,\ 1)$, $(\lambda_{\sigma_r},\ \lambda_{\sigma_r^{c\bar{c}}},\ \lambda_{\sigma^{\text{VM}}}) = (20,\ 25,\ 10)$ and $(\lambda_{\text{IR}},\ \lambda_{\text{pos}},\ \lambda_{\text{UV}}) = (1,\ 10000,\ 1000)$. Their values are determined through gradient-based loss-weight calibration, as described in the next subsection.}

The loss is minimized using the AdamW optimizer~\cite{loshchilov2018decoupled} with a cosine-annealing learning-rate schedule~\cite{DBLP:journals/corr/LoshchilovH16a}, ensuring stable and high-precision convergence. Training was run for 3000 epochs on a GPU, with the learning rate initialized at $10^{-3}$.

\subsubsection*{{A3. Gradient-based loss-weight calibration}}
{Our PINN objective combines multiple loss terms with different physical dimensions, numerical scales, and gradient magnitudes, which often leads to ill-conditioned optimization~\cite{doi:10.1137/20M1318043,WANG2024113112}. Inspired by gradient-normalization ideas from multi-task learning and PINNs~\cite{DBLP:journals/corr/abs-1711-02257,WANG2024113112}, we calibrate the loss weights using a gradient-based procedure before production training. The total loss is
\begin{equation}
    \mathcal{L}_{\text{total}} = \sum_i w_i \mathcal{L}_i,
\end{equation}
where $i$ indexes the ciBK residual, experimental data, Fourier-positivity, and physical-constraint losses.  
The weights $w_i$ are determined from the unweighted gradient magnitudes. We define the $L_2$ norm
\begin{equation}
    G_i = \lVert \nabla_{\theta}\mathcal{L}_i \rVert,
\end{equation}
computed during a warm-up stage, take the median $G^{\rm med} = \mathrm{median}(G_i)$ as a reference, and set
\begin{equation}
    w_i = \frac{G^{\rm med}}{G_i}.
\end{equation}
This brings all weighted gradient norms $w_i G_i$ to the same order of magnitude, preventing any single term from dominating the optimization solely because of its numerical scale. The weights are then kept fixed for the remainder of training.
}

\subsection{Theoretical Formalism for the Observables and Supplementary Results}

This section first summarizes the theoretical formalism used to compute the observables, and then presents comprehensive results for the reduced cross section $\sigma_r$ over all available HERA kinematic range. Additionally, we evaluate the longitudinal structure function $F_L$ as a cross-check, complementing the representative comparisons shown in the main text.

\subsubsection*{B1. Reduced cross sections and  longitudinal structure function: $\sigma_r$, $F_L$ and $\sigma_r^{c\bar{c}}$}
The DIS cross section is commonly written in terms of the reduced cross section,
\begin{align} \label{eq:sigma_r}
    \sigma_r(y,x_B,Q^2) =&\ F_2(x_{B}, Q^2)\notag \\
    &- {y^2 \over 1+(1-y)^2}F_L(x_B,Q^2) ,
\end{align}
where $y=Q^2/(s x_B)$ is the inelasticity, $x_B$ the Bjorken variable, $Q^2$ the photon virtuality, $\sqrt{s}$ the lepton--proton center-of-mass energy.

The structure functions $F_2$ and $F_L$ are related to the transverse (T) and longitudinal (L) virtual-photon proton cross sections via
\begin{align}\label{F2}    
    F_2 &= {Q^2 \over 4\pi^2 \alpha_{em}} \paren{\sigma_T^{\gamma^*p} + \sigma_L^{\gamma^* p}},\\
\label{eq:structure_functions}    F_L &= {Q^2 \over 4\pi^2 \alpha_{em}} \sigma_L^{\gamma^*p}.
\end{align}
Here  $\alpha_{em}$ is the electromagnetic coupling.

In the leading-order eikonal approximation within the color dipole picture~\cite{Kovchegov:2012mbw}, the $\gamma^{*}p$ interaction proceeds in two steps. First, the virtual photon fluctuates into a quark--antiquark pair, $\gamma^{*}\to q\bar q$, described by the light-cone wave function $\Psi^{\gamma^{*}\to q\bar q}$. The resulting color dipole then scatters off the proton. The corresponding $\gamma^{*}p$ cross sections for transverse and longitudinal photon polarizations are
\begin{align} 
    \sigma_{T,L}^{\gamma^*p} = 2\sum_f \int d^2 \mathbf{b} d^2 \mathbf{r} {dz \over 4\pi} |\Psi_{T,L}^{\gamma^* \to q\bar{q}}\paren{\mathbf{r},Q^2,z}|^2 \notag \\
    \times N(\mathbf{r}, \mathbf{b},x_B),
\end{align}
Here $\mathbf{r}$ is the transverse size of the $q\bar q$ dipole, $z$ is the fraction of the photon longitudinal momentum carried by the quark, $\mathbf{b}$ is the impact parameter, and $N(\mathbf{r}, \mathbf{b},x_B)$ denotes the dipole--proton scattering amplitude. The sum runs over the light quark flavors and the charm quark.

In this work, we neglect the impact-parameter dependence of the dipole--proton scattering amplitude. Accordingly, we replace the impact-parameter integral by a constant effective proton area, $\int d^{2}\mathbf{b}\;\longrightarrow\;{\sigma_{0}}/{2}\,,$ where $\sigma_{0}/2$ is the proton transverse area.

The squared photon light-cone wave functions, summed over quark helicities and, for the transverse case, averaged over the two transverse photon polarizations, are~\cite{Kovchegov:2012mbw}
\begin{align}
    |\Psi_L(r,Q^2,z)|^2 =&\ {8N_c \over \pi} \alpha_{em} e_q^2 Q^2 z^2 (1-z)^2 K_0^2(\varepsilon r), \\
    |\Psi_T(r,Q^2,z)|^2 =&\ {2N_c \over \pi} \alpha_{em} e_q^2 \Big\{ [z^2 + (1-z)^2]\varepsilon^2 K_1^2(\varepsilon r) \notag \\ 
    &\ + m_f^2 K_0^2(\varepsilon r) \Big\},
\end{align}
where $r = |\mathbf{r}|$, $\varepsilon^2 = z(1-z)Q^2 + m_f^2$, $e_f$ is the fractional charge of quark flavor $f$, and $K_0$ and $K_1$ are modified Bessel functions of the second kind of order zero and one, respectively. For the light and charm quarks we use the effective masses $m_{\text{light}}=0.14~\mathrm{GeV}$ and $m_c=1.4~\mathrm{GeV}$, respectively.

In practice, when computing $\sigma^{\gamma^{*}p}_{T,L}$, we apply the standard kinematical shift and introduce the modified Bjorken variable
\begin{equation}
    \tilde{x} = x_B \left(1 + \frac{4m_f^2}{Q^2} \right)\, .
\end{equation}
 for each quark flavor.

\subsubsection*{B2. Diffractive Vector-Meson Production}
In exclusive vector-meson production at small Bjorken-\(x\), the relevant longitudinal momentum fraction is set by the hard scale \(Q^2+M_V^2\) and the photon–proton center-of-mass energy \(W\),
\begin{equation}
x \equiv \frac{Q^2+M_V^2}{W^2+Q^2}\,.
\end{equation}
The diffractive scattering amplitude for \(\gamma^* p \to Vp\) reads~\cite{Kowalski:2006hc}
\begin{align}
    \mathcal{A}_{T,L}^{\gamma^* p \to Vp}(x,Q^2,\mathbf{\Delta}) = i &\int d^2\mathbf{r} \int_0^1 {dz \over 4\pi} \int d^2 \mathbf{b} \paren{\Psi_V^* \Psi}_{T,L} \notag \\
    &\times e^{-i[\mathbf{b} - \paren{1-z} \mathbf{r}]\cdot \mathbf{\Delta}}\  {d\sigma_{q \bar{q}} \over d^2 \mathbf{b} },
\end{align}
where \(\boldsymbol{\Delta}\) is the momentum transfer, related to the Mandelstam variable \(t\) by \(t=-\Delta^2\), and \(d\sigma_{q\bar q}/d^2\mathbf{b}\) is the dipole–proton cross section differential in the impact parameter \(\mathbf{b}\). 
Assuming impact-parameter independence, the \(b\)-integrated dipole cross section becomes
\begin{equation}
\sigma_{q\bar q}(x,r)\equiv \int d^2\mathbf{b}\,\frac{d\sigma_{q\bar q}}{d^2\mathbf{b}}
= \sigma_0\,N(x,r)\,.
\end{equation}

The transverse and longitudinal photon and vector-meson overlap functions are given by~\cite{Kowalski:2006hc}
\begin{align}
    \paren{\Psi_V^* \Psi}_{T} =& \hat{e}_f e {N_c \over \pi z(1-z)}\Big\{ m_f^2 K_0(\varepsilon r)\phi_T(r,z)\notag \\
        &- [z^2 + (1-z)^2]\varepsilon K_1(\varepsilon r) \partial_r\phi_T(r,z) \Big \},
\end{align}

\begin{align}
    \paren{\Psi_V^* \Psi}_{L} =& \hat{e}_f e{N_c \over \pi} 2 Q z (1-z) K_0(\varepsilon r)\Big[M_V\phi_L(r,z)\notag \\
    &+ \delta {m_f^2 - \nabla_r^2 \over M_V z (1-z)}\phi_L(r,z) \Big],
\end{align}
where \(\hat e_f\) is the fractional electric charge of the charm
quark,  \(e=\sqrt{4\pi\alpha_{\rm em}}\), and \(M_V=3.097~\mathrm{GeV}\) is the \(J/\psi\) mass. The scalar functions \(\phi_{T,L}(r,z)\) encode the meson wave function, in this work we use the boosted-Gaussian parametrization for \(J/\psi\)~\cite{Kowalski:2006hc}. 

The \(t\)-differential cross section for exclusive vector-meson production is given by~\cite{Kowalski:2006hc}
\begin{equation}\label{eq:dsigma_dt}
\frac{d\sigma_{T,L}^{\gamma^* p \to Vp}}{dt}
= \frac{R_g^2}{16\pi}\,
\left|\mathcal{A}_{T,L}^{\gamma^* p \to Vp}\right|^2\,
\left(1+\beta^2\right),
\end{equation}
where \(\beta\equiv \mathrm{Re}\,\mathcal{A}/\mathrm{Im}\,\mathcal{A}\) accounts for the real part of the scattering amplitude contribution. It can be calculated from the effective small-\(x\) exponent of the amplitude,
\begin{equation}
\beta = \tan\!\left(\frac{\pi \lambda}{2}\right),
\qquad
\lambda
= \frac{\partial \ln\!\left(\mathcal{A}_{T,L}^{\gamma^* p \to Vp}\right)}{\partial \ln(1/x)}\,.
\end{equation}
The skewness (off-forward) correction is encoded in \(R_g\), which in the Shuvaev prescription reads~\cite{Shuvaev:1999ce}
\begin{equation}
R_g
= \frac{2^{\,2\lambda+3}}{\sqrt{\pi}}\,
\frac{\Gamma\!\left(\lambda+\tfrac{5}{2}\right)}{\Gamma\!\left(\lambda+4\right)}\,.
\end{equation}

Integrating Eq.~\eqref{eq:dsigma_dt} over \(t\) yields the total cross section,
\begin{align}\label{VM_cs_total}
\sigma_{T,L}^{\gamma^* p \to Vp}
&= \int dt\, \frac{d\sigma_{T,L}^{\gamma^* p \to Vp}}{dt}\,,
\\
\sigma_{\mathrm{total}}^{\gamma^* p \to Vp}
&= \sigma_{T}^{\gamma^* p \to Vp} + \sigma_{L}^{\gamma^* p \to Vp}\,.
\end{align}
In the present study we focus on \(J/\psi\) photoproduction with \(Q^2=0\), for which only the transverse contribution is relevant.

\subsubsection*{B3. Reduced cross sections and longitudinal structure function at all energies}

\begin{figure*}
    \centering
    \includegraphics[width=0.9\textwidth]{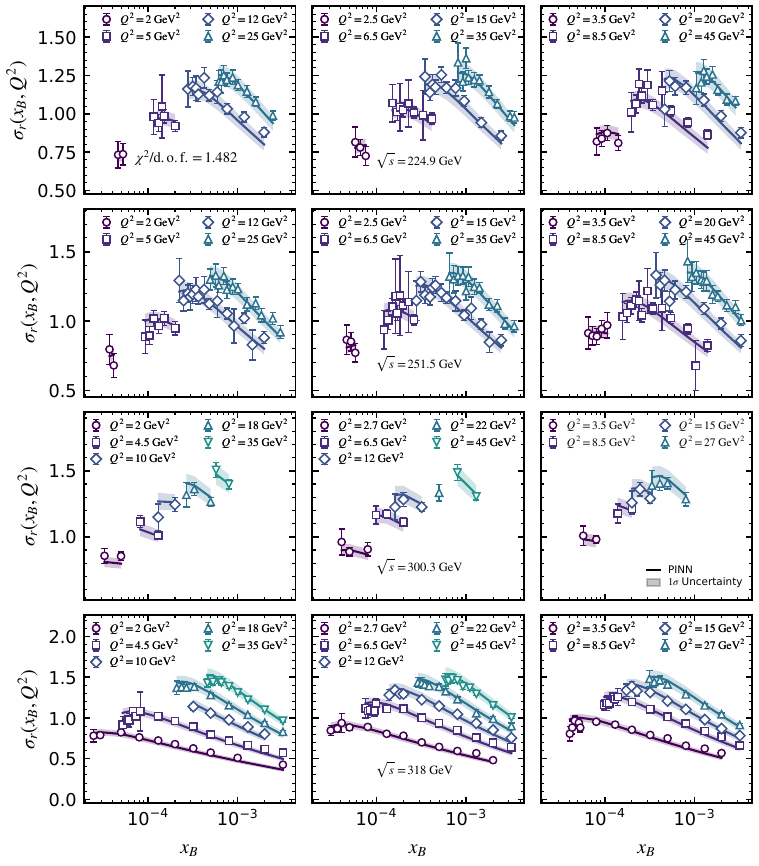}
    \caption{The reduced total cross section $\sigma_r(x_B,Q^2)$ compared with combined HERA $e^+p$ data~\cite{H1:2015ubc,H1:2009pze} at all available center-of-mass energies, $\sqrt{s}=224.9,\ 251.5,\ 300.3,\ 318\ \mathrm{GeV}.$ The quoted $\chi^2/\mathrm{d.o.f.}$ uses the solid curve, and it includes all available energies.}
    \label{fig:sigma_r_full}
\end{figure*}

This section presents supplementary comparisons for the inclusive reduced cross section $\sigma_r$ at all available HERA center-of-mass energies, and for the longitudinal structure function $F_L$ as a cross-check. Figure~\ref{fig:sigma_r_full} compares the PINN predictions with the combined HERA measurements of the inclusive reduced total cross section \(\sigma_r\) at the center-of-mass energies \(\sqrt{s}=224.9,\,251.5,\,300.3,\) and \(318~\mathrm{GeV}\). The good agreement over the measured kinematic range provides a global validation of the extracted dipole amplitude in inclusive DIS.

\begin{figure*}
    \centering
    \includegraphics[width=0.9\textwidth]{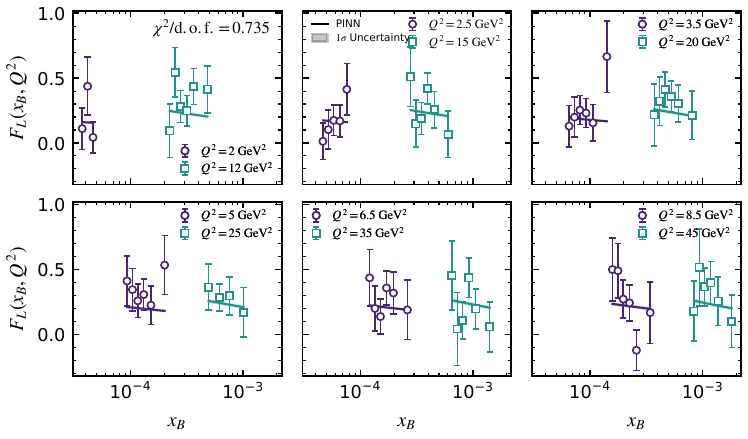}
    \caption{The longitudinal structure function $F_L(x_B, Q^2)$ compared with HERA $e^+p$ data~\cite{H1:2013ktq} at different $Q^2$ at all available center-of-mass energies.}
    \label{fig:FL}
\end{figure*}

Figure~\ref{fig:FL} compares the PINN predictions with the $F_L$ measurements. The significance of this cross-check lies in isolating the longitudinal photon contribution through Eq.~\eqref{eq:structure_functions}. Since $\sigma_r$ depends on both $F_2$ and $F_L$ (see Eq.~\eqref{eq:sigma_r}), these constraints are not fully independent, and the $F_L$ comparison provides a nontrivial consistency check of the extracted universal amplitude. The resulting uncertainty band for $F_L$ is comparatively narrow in our fit, consistent with the indirect constraint provided by $\sigma_r$. The agreement with the data supports the robustness of the extracted $N(r,Y)$ across observables that emphasize different photon polarizations. 

\end{document}